\documentclass[a4, 11pt]{article}
\usepackage[T1]{fontenc}

\usepackage{amssymb}        
\usepackage{latexsym}
\usepackage{times}

\usepackage{graphicx}

\author{}

\date{}

\title{Popular $b$-Matchings}

\author{Katarzyna Paluch \\
Institute of Computer Science, Wroc{\l}aw University }

\newcommand{\dowod}{\noindent{\bf Proof.~}}
\newcommand{\koniec}{\hfill $\Box$\\[.1ex]}

\newtheorem{lemma}{Lemma}
\newtheorem{theorem}{Theorem}

\newtheorem{definition}{Definition}

\begin{document}

        \maketitle
\thispagestyle{empty}
\section{Introduction}
In the paper we study popular $b$-matchings, which in other words are popular many-to-many matchings. The problem can be best described in graph terms:
We are given a bipartite graph $G=(A \cup H, E)$, a capacity function on vertices $b: A \cup H \rightarrow N$ and a rank function  on edges $r: E \rightarrow N$. $A$ stands for the set of agents and $H$ for the set of houses.
Each member of a set of agents $a \in A$
has a  preference list $P_a$ of a subset $H_a$ of houses $H$. For $a \in A$ and $h \in H$ edge $e=(a,h)$ belongs to $E$ 
iff $h$ is on $P_a$ and $r((a,h))=i$ reads that $h$ belongs to (one of) $a$'s $i$th choices. 
We say that $a$ {\bf prefers} $h_1$ to $h_2$ (or {\bf ranks $h_1$ higher than $h_2$}) if $r((a,h_1)) < r((a,h_2))$.
If $r(e_1)<r(e_2)$ we say that $e_1$ has a {\bf higher rank} than $e_2$.
If there exist $a\in A$ and $h_1, h_2 \in H_a, h_1 \neq h_2$ such that $r(e_1=(a,h_1))=r(e_2=(a,h_2))$, then we say
that $e_1, e_2$ belong to a {\bf tie} and graph $G$ contains ties. Otherwise we say that $G$ does not contain ties.
 A {\bf $b$-matching} $M$ of $G$ is such a subset of edges that $deg_M(v) \leq b(v)$ for every $v \in A \cup H$,
meaning that every vertex $v$ has at most $b(v)$ edges of $M$ incident with it.
Let $r$ denote the greatest rank (i.e. the largest number) given to any edge of $E$. 
With each agent $a$ and each $b$-matching $M$ we associate a {\bf signature} denoted as $sig_M(a)$, which is an $r$-tuple $(x_1,x_2, \ldots, x_r)$
such that $x_i$ ($1 \leq i \leq r$) is equal to the number of edges of rank $i$ matched to $a$ in a $b$-matching $M$.
We introduce a {\bf lexicographic order $\succ$ on signatures} as follows. We will say that $ (x_1,x_2, \ldots, x_r) \succ
(y_1,y_2, \ldots, y_r)$ if there exists $j$ such that $1 \leq j\leq r$ and for each $1 \leq i \leq j-1$ there is $x_i=y_i$ and $x_j > y_j$.
We say that an agent $a$ {\bf prefers} $b$-matching $M'$ to $M$ 
if $sig_{M'}(a) \succ sig_M(a)$. $M'$ is {\bf more popular} than $M$, denoted by $M' \succ M$, if the number of agents that prefer $M'$ to $M$ exceeds the number of agents that prefer $M$ to $M'$.   

\begin{definition}
A $b$-matching $M$ is {\bf popular} if  there exists no $b$-matching $M'$ that is more popular than $M$.
The {\bf popular $b$-matching problem} is to determine if a given triple $(G,b,r)$ admits a popular $b$-matching and
find one if it exists.

\end{definition}

\noindent {\bf Previous work} The notion of popularity was first introduced by Gardenfors \cite{Gar} in the one-to-one and two-sided context, where two-sided means that both agents and houses express their preferences over the other side and a matching $M$
is popular if there is no other matching $M'$ such that more participants (i.e. agents plus houses) prefer $M'$ to $M$ than $M$ to $M'$. (He used the term of a {\em majority $1$assignment}.)  He proved that every stable matching is a popular matching if there are no ties. \\
One-sided popular matchings were first studied in the one-to-one setting by Abraham et al. in \cite{Abr1}. They proved that a popular matching needn't exist and decribed fast polynomial algorithms
to compute a popular matching, if it exists. 
The dynamic scenario, in which agents and houses can come and leave the graph and agents can change their preference lists, is examined in \cite{Abr2}. Mestre \cite{Mestre} considered a version in which every agent has an associated weight, reflecting an agent's priority. Manlove and Sng in \cite{Esa} extended an algorithm from \cite{Abr1}
to the one-to-many setting (notice that this not equivalent to the many-to-one setting.) In \cite{Sng}, in turn, the version, in which agents have weights, houses have arbitrary capacities, every agent has capacity one and there are no ties, is examined. \\
Mahdian \cite{Mahd} showed for the one-to-one case without ties that a popular matching exists with high probability
when  preference lists are random, and  the number of houses is a small multiplicative factor larger than the number of agents. Instances when a popular matching does not exist were dealt with by McCutchen (\cite{McCutchen}), who defined
two notions of a matching that is, in some sense, as popular as possible, namely a leastunpopularity-
factor matching and a least-unpopularity-margin matching and  proved that computing either type of matching is $NP$-hard,
even if there are no ties,  by Huang et al. \cite{Huang} and by Kavitha et al. \cite{Mixed}.
Kavitha and Nasre \cite{Nasre} gave an  algorithm to 
compute an optimal popular matching for various interpretations of optimality.
McDermid and Irving \cite{Mcdermid} gave a characterisation
of the set of popular matchings for the one-to-one version without ties, which can be exploited to yield algorithms for related problems. 

\noindent{\bf Our contribution}
We provide a characterization of popular $b$-matchings and prove that the popular $b$-matching problem is $NP$-hard
 even when agents use only two ranks and have capacity at most $2$ and houses have capacity one. This in particular answers the question about
many-to-one popular matchings asked in \cite{Esa}. Next we modify the notion of popularity and consider so-called {\em
weakly popular $b$-matchings}. We give their characterization and show that finding a weakly popular $b$-matching or reporting that it does not exist is $NP$-hard even if all agents use at most three ranks, there are no ties and
houses have capacity one. We construct polynomial algorithms for the versions in which agents use two ranks.

\section{Characterization}
First we introduce some terminology and recall a few facts from the matching theory. \\
By a path $P$ we will mean a sequence of edges. Usually a path  $P$ will be denoted as $(v_1, v_2, \ldots, v_k)$,
where $v_1, \ldots, v_k$ are vertices from the graph, not necessarrily all different, and for each $i$ ($ 1 \leq i \leq k-1$) $(v_i, v_{i+1})\in E$  and no edge of $G$ occurs twice in $P$. We will  sometimes treat a path as a sequence of edges and sometimes as a set of edges. \\
If $M$ is a $b$-matching and $deg_M(v)<b(v)$ we will say that $v$ is {\bf unsaturated}, if $deg_M(v)=v(v)$ we say that $v$ is {\bf saturated}. If $b(v)=1$, then we will also use the terms {\bf matched} and {\bf unmatched} instead of saturated and unsaturated. If $e\in M$ we will call it an {\bf $M$-edge} and otherwise -- a {\bf non-$M$-edge}. 
By $M(v)$ we mean the set $\{v': (v,v') \in M\}$. A path is said to be {\bf alternating} (with respect to $M$) or {\bf $M$-alternating} if its edges are alternately $M$-edges and non-$M$-edges. An alternating path is said to be {\bf ($M$-)augmenting} if its end vertices are unsaturated and its first and last edges are non-$M$-edges.\\
For two sets $Z_1, Z_2 $ $Z_1 \oplus Z_2$ is defined as $(Z_1 \setminus Z_2) \cup (Z_2 \setminus Z_1)$. 
If $M$ is a $b$-matching and $P$ is an alternating with respect to $M$ path such that its beginning edge $(v_1,v_2)$
is an $M$-edge or $v_1$ is unsaturated and its ending edge $(v_{k-1},v_k)$ is an $M$-edge or $v_k$ is unsaturated, then $M \oplus P$ is a also a $b$-matching
and if $P$ is additionally augmenting, then $M \oplus P$ has more edges than $M$ and is said to have larger size or greater
cardinality than $M$. A $b$-matching of maximum size is called a {\bf maximum $b$-matching}.\\
We will also need a notion of an {\bf even path}: a path 
$(a_1,h_1,a_2,h_2, \ldots, v)$, where $v$ denotes either $h_k$ or $a_{k+1}$, is defined to be even and denoted $P_e(a_1,v)$ if it is alternating, $(a_1,h_1)$ is a non-$M$-edge and for each $i, 2 \leq i \leq k$ edges $(h_{i-1},a_i), (a_i, h_i)$ have the same rank.
If this path is written in the reverse order we denote it by $P^r_e(v,a_1)$.
By writing $(P_e(a,h),a')$ we mean a path that consists of an even path $P_e(a,h)$ and edge $(h,a')$.
 
\begin{theorem} \label{char}
A $b$-matching $M$ is popular iff graph $G$ does not contain a  path of one of the  following four types:
\begin{enumerate}
\item $(P_e(a_1,h_{k-1}), a_k, h_k, a_{k+1})$, where $(1)\ $ the path is alternating and $(2)\ $   $a_1$ is unsaturated or there exists an edge $e$ in $M(a)$ such that  $r(e)>r(a_1,h_1)$,  $(3)\ $ $r(h_{k-1},a_k)>r(a_k, h_k)$ and $(4)$ agents $a_1,a_k, a_{k+1}$
are pairwise different or $a_1 \neq a_k, a_1=a_{k+1}$ and $r(a_1,h_1)<r(a_1,h_k)$,

\item $(P_e(a_1,a),P^r_e(a, a'_1))$, where  $(1)\ $ $a_1$  is unsaturated or there exists an edge $e$ in $M(a_1)$  such that $r(e)>r(a_1,h_1)$, $(2)\ $ $a'_1$  is unsaturated or there exists an edge $e'$ in $M(a'_1)$  such that $r(e')>r(a'_1,h'_1)$ and $(3)$ agents $a_1, a, a'_1$ are pairwise different,

\item $P_e(a_1,h)$, $(1)\  $ $a_1$ is unsaturated or there exists an edge $e$ in $M(a)$ such that $r(e)>r(a_1,h_1)$ and $(2)\ $ $h$ is unsaturated,

\item $(P_e(a_1,h_k), a_1)$,  where the path is alternating and $r(h_k,a_1)>r(a_1,h_1)$.

\end{enumerate} 
\end{theorem}
\dowod
Suppose  the graph contains a path $P_1$ of the first type.  If $a_1$ is saturated let $P'_1=P_1 \cup e$ (notice that $e \notin P_1$),
otherwise let $P'_1=P_1$.
$M'=M \oplus P'_1$ is a $b$-matching such that
$sig_{M'}(a_1) \succ sig_M(a_1), sig_{M'}(a_k) \succ sig_M(a_k), sig{M}(a_{k+1}) \succ sig_M'(a_{k+1}) $ and for each $a$ different from $a_1, a_k, a_{k+1}$
we have $sig_M'(a)=sig_M(a)$. Therefore $M'$ is more popular than $M$.

If the graph contains a path of the second type, we proceed analogously. 

Suppose the graph contains a path $P_3$ of the third type. If $a_1$ is saturated let $P'_3=P_3 \cup e$,
otherwise let $P'_3=P_3$. $M'=M \oplus P'_3$ is a $b$-matching such that
$sig_{M'}(a_1) \succ sig_M(a_1)$ and for each $a$ different from $a_1$
we have $sig_M'(a)=sig_M(a)$. Therefore $M'$ is more popular than $M$.

Suppose now that there exists a $b$-matching $M'$ which is more popular than $M$.
This means that the set $A_1$ of agents who prefer $M'$ to $M$ outnumbers the set $A_2$ of agents who prefer $M$ to $M'$.

For each $a \in A_1$ we build a path $P_a$ in the following way. 
We will use only edges of $M\oplus M'$. We start with an edge $(a,h_1) \in M'\setminus M$
having the highest possible rank (i.e. lowest possible number). Now assume that our so far built path $P_a$ ends with $h_i$.
 If $h_i$ is unsaturated in $M$, we end. Otherwise we consider edges $(h_i, a_i)$ belonging
to $M \setminus M'$ and not already used by other paths $P_a'$ ($a' \in A_1$). (The set of such unused edges is nonempty as $h_i$ is saturated in $M$ and thus there are at least as many $M$-edges as $M'$-edges incident with $h_i$ and each time we arrive at $h_i$ while building some path $P_a$ ($a \in A_1$) we use one $M$-edge and one $M'$-edge.) If among these edges, there is such one that $a_i \in A_1$ we add it to $P_a$ and stop. Otherwise if there is such one that $a_i \in A_2$ we add it to $P_a$ and stop. Otherwise we add  any unused edge $(h_i,a_i)$ to $P_a$. $a_i$ has the same signature in $M$ and in $M'$.
Therefore there exists an  edge $(a_i,h_{i+1}) \in M' \setminus M$ having the same rank as $(h_i, a_i)$ and there 
exists an unused edge of this kind because the number of edges in $M$ incident with $a_i$ is the same as the number 
of edges in $M'$ incident with $a_i$ (as $a_i$ has the same signature in both $b$-matchings), we add this edge to $P_a$.

Clearly we stop building 
$P_a$ at some point because we either arrive at an unsaturated vertex $h \in H$ or at a vertex of $A_1 \cup A_2$, which may be $a$ itself. 
Suppose there exists $a \in A_1$ such that $P_a$ ends on $a$. Since we have started from an edge $e$ of $M'(a)\setminus M(a)$ having the highest rank, the ending edge of $P_a$ must have a lower rank than $e$, hence $P_a$ is a path of the fourth type. 
If there exists a path $P_a$ ending on an unsaturated vertex, it is of type three. If there exists
$a \in A_1$ such that $P_a$ ends on $a_1 \in A_1, a_1 \neq a$, then let $e=(h',a_1)$ denote the ending edge of $P_a$.
Since $a_1 \in A_1$ the edge of $M(a_1)\setminus M'(a_1)$ having the highest rank, let us call it $e'$ has a higher rank than $e$. Suppose $e'=(a_1,h)$. If there exists an edge $e_3 = (h,a) \in M \setminus M'$, path $P_a \cup e' \cup e_3$
forms a path of type $(1)$. Otherwise there exists an edge $e_3=(h,a_2)\in M \setminus M'$, where $a_2 \neq a$ and
of course $a_2 \neq a_1$ and path $P_a \cup e' \cup e_3$ also
forms a path of type $(1)$.
If none of the above paths exists, each $P_a$ ends on some  agent $a_2 \in A_2$.
Because $A_2$ outnumbers $A_1$ there exist $a_1,a'_1\in A_1, \ a_1 \neq a'_1$ and $a_2 \in A_2$ such that $P_{a_1}$
and $P_{a'_1}$ both end on $a_2$. These paths are edge-disjoint and together form a path of type $(2)$. 
\koniec

\begin{theorem}
The problem of deciding whether a given triple $(G,b,r)$ has a popular $b$-matching is $NP$-hard, even if
all edges are of one of two ranks, each applicant $a \in A$ has capacity at most $2$ and each house $h \in H$ has capacity $1$.
\end{theorem}

\begin{figure} 
\centering{\includegraphics[scale=0.6]{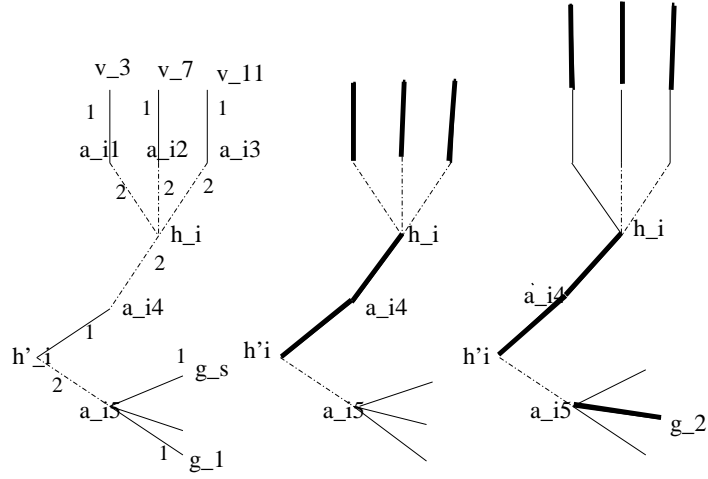}}
\caption{\scriptsize  Solid lines denote edges of rank $1$ and dotted lines edges of rank $2$. In the second and third
part of the figure thick lines indicate $M$-edges. The second part describes the situation when $T_i=\{3,7,11\}$ belongs
to an exact $3$-cover $J$ and the third part the situation when $i \notin J$.
} \label{gadzet1}
\end{figure}

\dowod
We prove the theorem by showing a polynomial reduction of the exact $3$-cover problem to the popular $b$-matching problem. In the {\em exact $3$-cover problem} we have a finite set $K$ and a family $\Sigma = \{T_i: i=1,2, \ldots, m\}$ of subsets of $K$ such that $|T_i|=3$ for $1 \leq i \leq m$. We want to establish if there exists $J \subseteq \{1,2, \ldots, m\}$ such that $\{S_j\}_{j \in J}$ forms a partition of $K$.
The exact $3$-cover problem is NP-complete even when each element of $K$ belongs to either two o three sets of $\Sigma$
and the underlining graph is planar \cite{Dyer}, \cite{Cor}.
Given an instance of the exact $3$-cover problem, we construct the following graph $G=(A \cup H, E)$. We have a vertex
$v_k \in H$ for each element $k \in K$, with $b(v_k)=1$. For each set $T_i$ we have $5$ vertices $a_{i_1}, a_{i_2}, a_{i_3}, a_{i_4}, a_{i_5} \in A$ with $b(a_{i_j})=1$ for $j\neq 4$ and $b(a_{i_4})=2$ and $2$ vertices $h_i, h'_i \in H$ with $b(h_i)=b(h'_i)=1$. Additionally we have $m-|K|/3$ vertices $g_1, g_2, \ldots g_{m-|K|/3} \in H$
having capacity $1$. The subgraph of $G$ corresponding to a set $T_i$ is shown in Figure\ref{gadzet1}.  Each vertex $v_k$ is connected by a rank one edge with 
one of vertices $a_{i_1}, a_{i_2}, a_{i_3}$ for each $i$ such that $k \in T_i$. Each vertex $a_{i_5}$ is connected
with each vertex of $\{g_1, g_2, \ldots g_{m-|K|/3}\}$ by a rank one edge.

Suppose there exists $J$ which is an exact $3$-cover of $K$. We build a popular $b$-matching $M$ as follows. For each $T_j=\{j_1, j_2, j_3\}$ such that $j \in J$ we add edges $(v_{j_1}, a_{j_1}),
(v_{j_2}, a_{j_2}), (v_{j_3}, a_{j_3})$ to a $b$-matching $M$. For each $i=1,2,\ldots, m$ we add edges $(a_{i_4}, h_i),
(a_{i_4}, h'_i)$ to $M$ and for each $i \notin J$ we add an edge $(a_{i_5},g)$, where $g$ is some vertex of $\{g_1, g_2, \ldots g_{m-|K|/3}\}$. We claim that $M$ is popular. Since all vertices of $H$ are saturated,
there does not exist a path of type $3$ from Theorem \ref{char}. Each vertex $a_{i_4}$ is saturated and matched via one rank one edge
and one rank two edge.  If some vertex $a_{i_j}$ is unsaturated, then it has no $M$-edges incident with it. Therefore it cannot belong to a path of type $4$. If it is saturated, then it is matched via a rank one edge and cannot
belong to a path of type $4$. If $a_{i_1}$ is unsaturated and its rank one edge is $(a_{i_1}, v_1)$,
then $v_1$ is matched via a rank one edge to some $a_{i'_j}$ where $j\in \{1,2,3\}$ and $a_{i'_j}$ has only one rank one edge incident with it. Thus $a_{i_1}$ cannot be a beginning of a path of type $1$. If $a_{i_1}$ has no rank one $M$-edge incident with it, then
$i \notin J$ and $a_{i_5}$ is saturated. Therefore path $(a_{i_1},h_i,a_{i_4},h'_i,a_{i_5})$ is not a path of type $2$. If $a_{i_5}$ is unmatched, then $a_{i_1}, a_{i_2}, a_{i_3}$ are saturated and thus path $(a_{i_5},h'_i,a_{i_4},h_i,a_{i_1})$ is not a path of type $2$. Also unmatched $a_{i_5}$ does not belong to any path of type $1$.

Next we show, that if there is no exact $3$-cover of $K$, then $G$ has no popular $b$-matching.
First let us notice that if a $b$-matching $M$ of $G$ is popular, then all vertices of  $H$  are saturated in $M$.
Next we can see that each edge $(a_{i_4}, h'_i)$ belongs to every popular $b$-matching of $G$, for if it does not belong to some $b$-matching $M$, then path $(a_{i_4}, h'_i, a_{i_5}, g_1, a_{i'_5})$, where $a_{i'_5}$ is a vertex matched to $g_1$, is of type $1$ from  Theorem \ref{char}.
Also in every popular $b$-matching $M$ $a_{i_4}$ must be saturated. For if edge $(a_{i_4}, h_i)$ does not belong to
$M$, then $h_i$ is matched to a vertex of $\{a_{i_1}, a_{i_2}, a_{i_3}\}$, say $a_{i_1}$. Then path $(a_{i_4}, h_i,
a_{i_1}, v_s, a_{i'_j})$ , where $v_s$ is a neighbour of $a_{i_1}$ and $a_{i'_j}$ is matched to $v_s$, is a path
of type $1$ from Theorem \ref{char}.
If there is no exact $3$-cover of $K$, then for some $i$ only two or one vertices of $\{a_{i_1}, a_{i_2}, a_{i_3}\}$
will be matched via  a rank one edge. The number of such $i$ clearly surpasses $m-|K|/3$. Suppose that $a_{i_1}$ is not matched via a rank one edge and $a_{i_2}, a_{i_3}$
are. Then if path $(a_{i_1}, h_i, a_{i_4}, h'_i, a_{i_5})$ is not to be a path of type $2$, $a_{i_5}$ must be saturated. But only $m-|K|/3$ vertices of $\{a_{i_5}: i=1,2, \ldots m\}$ can be saturated.  

\koniec

\section{Weakly popular $b$-matchings}
Instead of just checking whether an agent prefers one $b$-matching to the other, we might also want to assess how much
he/she prefers one to the other. 

Before defining a loss/gain factor of an agent we introduce the following. 
For a set $E'=\{e_1, e_2, \ldots, e_k\}$ of edges incident with $a$, let $s(E')$ denote a $d$-tuple $(x_1,x_2, \ldots, x_d)$ such that $x_i=r(e_{j_i})$ if $1 \leq i \leq k$ and $x_i=r+1$ if $i>k$, where $e_{j_1}, e_{j_2}, \ldots, e_{j_k}$ are edges of $E'$ ordered so that $r(e_{i_j}) \leq r(e_{i_{j+1}})$ for $ 1 \leq j \leq k-1$ and $d=max \{deg(a): a\in A\}$.  For two $d$-tuples $d_1=(x_1, \ldots, x_d), \ d_2=(y_1, y_2, \ldots, y_d)$ we define their difference as $d_1-d_2= \Sigma_{i=1}^d signum(x_i-y_i)$, where $signum(x)=1, 0,$ or $-1$ if correspondingly $x>0, x=0$ or $x<0$.   
Now we define a {\em loss/gain factor} of agent $a$ with respect to $b$-matchings $M,M'$
as $f_a(M,M')= s(M\setminus M \cap M')- s(M' \setminus M \cap M')$.
Also we define a loss/gain factor for two $b$-matchings: $f(M,M')= \Sigma_{a\in A} f_a(M,M')$.
We will say that a $b$-matching $M$ is {\bf more weakly popular} than a $b$-matching $M'$ if $f(M,M')>0$ and a $b$-matching $M'$ will be called {\bf weakly popular} if there does not exist a $b$-matching $M'$ that is more weakly 
popular than $M$.

\begin{theorem} \label{char1}
A $b$-matching $M$ is weakly popular iff graph $G$ does not contain a path of type $(1),(3)$ or $(4)$ from Theorem \ref{char}.
\end{theorem}
The proof is similar to that of Theorem \ref{char}.

\begin{theorem}
The problem of deciding whether a given triple $(G,b,r)$ admits a weakly popular $b$-matching is $NP$-hard, even if
all edges are of one of three ranks, each applicant $a \in A$ has capacity at most $3$, each house $h \in H$ has capacity $1$ and there are no ties.
\end{theorem}

\begin{figure} 
\centering{\includegraphics[scale=0.5]{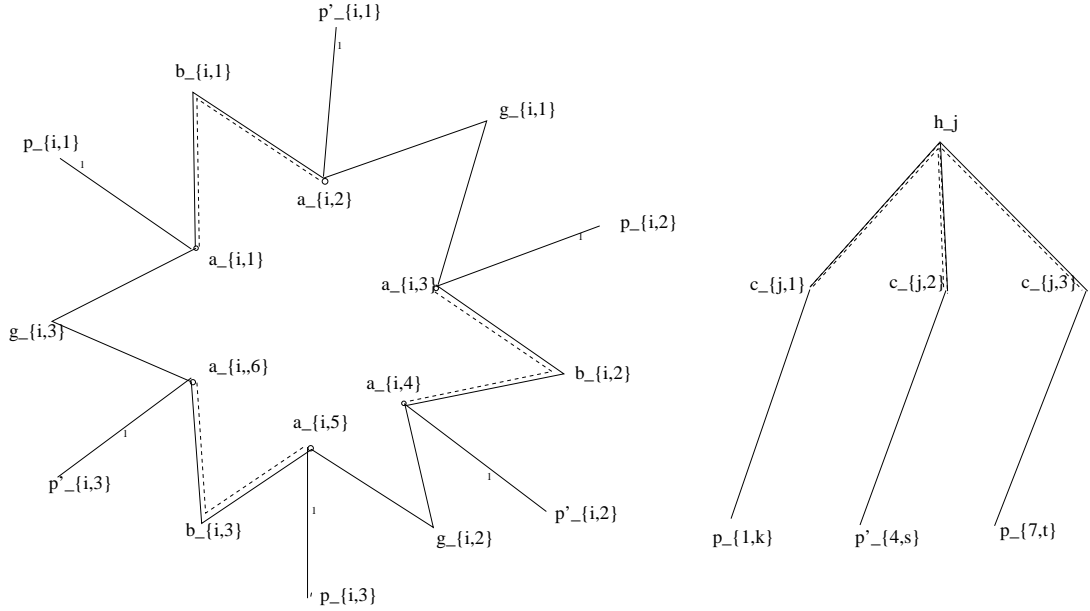}}
\caption{\scriptsize A subgraph corresponding to variable $p_i$ such that $r(i)=3$ and a subgraph corresponding to
the exemplary $1$st clause.
} \label{turbina}
\end{figure}

\dowod
We will reduce the $3$-SAT problem to the Weakly Popular $b$-matching problem.
assume we have a formula $F$  that uses $k$ variables $p_1, p_2, \ldots, p_k$ and has the form: $(q_{1,1} \vee q_{1,2} \vee q_{1,3})\wedge \ldots  \wedge (q_{n,1} \vee q_{n,2} \vee q_{n,3})$, where each $q_{j_1,j_2}$ ($1\leq j_1 \leq n,
\ 1 \leq j_2 \leq 3)$ represents either $p_i$ or $\overline{p_i}$ (meaning $not\ p_i$) for some $i$. For each $p_i$ let $r(p_i)$ denote the number of times  $p_i$  occurs in $F$ and $r'(p_i)$ -- the number of times $\overline{p_i}$ occurs in $F$.
Let $r(i)=\max\{r(p_i), r'(p_i)\}$. \\
We construct the following graph $G=(A \cup H, E)$.  For each variable $p_i$ we have the following vertices that will belong to $H$: $p_{i,1}, p_{i,2}, \ldots, p_{i,{r(i)}}$ and  $p'_{i,1}, \ldots,  p'_{i,{r(i)}}$ and $b_{i,1}, \ldots, b_{i,r(i)}$ and $g_{i,1}, \ldots, g_{i,r(i)}$. For each of the $n$ clauses we have one additional house $h_i$.
For each variable $p_i$ we have $2r(i)$ agents $a_{i,1}, \ldots, a_{i,2r(i)}$, each of capacity $3$.  For each odd $j$ agent $a_{i,j}$ is connected via a rank one edge with $p_{i,(j+1)/2}$ and for each even $j$ agent $a_{i,j}$ is connected via a rank one edge with $p'_{i,{j/2}}$. Agents $a_{i,j},a_{i,j+1}$ for each odd $j$ are connected via rank two edges with $b_{i,(j+1)/2}$ for each odd $j$ and agents $a_{i,j},a_{i,(j+1) \ mod\  2r(i)}$ for each even $j$ are connected via rank three edges to $g_{i,j/2}$ for each even $j$.
For each clause we have $3$ agents: $c_{i,1}, c_{i,2}, c_{i,3}$, each of capacity $2$. All of them are connected via a rank two edge with $h_i$.
If the $1$st clause is of the form, say $(p_1 \vee not\ p_4 \vee p_7)$, then $c_{1,1}$ is connected via a rank one edge with $p_{1,j}$ for some $j$, $c_{1,2}$ with $p'_{4,j}$ for some $j$ and $c_{1,3}$ with $p_{1,j}$ for some $j$. No vertex
$p_{i,j}$ or $p'_{i,j}$ is connected to two clause vertices. 
The construction is illustrated in Figure\ref{turbina}.

If $F$ is true when $p_i=f(p_i)$ ($1 \leq i \leq k$) for some function $f: \{p_1, \ldots, p_k\} \rightarrow \{0,1\}$,
we build the following $b$-matching $M$. If $f(p_i)=1$ $(1 \leq i \leq k)$, we add to $M$ all rank one, two and  three  edges incident with agents
$a_{i,j}$ such that $j$ is even, otherwise $M$ will contain all rank one, two and three edges incident with agents
$a_{i,j}$ such that $j$ is odd. Since $F$ is true, in the first clause we have $f(p_1)=1 \vee f(p_4)=0 \vee f(p_7)=1$.
Suppose that $f(p_4)=0$, then we add to $M$ a rank one edge incident with $c_{1,2}$, we are able to do so, because at the moment all vertices $p'_{4,j}$ are unmatched. We also add to $M$ a rank two edge incident with $c_{1,2}$.
We proceed in this way for every clause. At the end for every rank one edge $(a,h)$ such that $a$ and $h$ are unsaturated we add it to $M$. One can check that thus built $M$ is weakly popular.

\begin{figure} 
\centering{\includegraphics[scale=0.5]{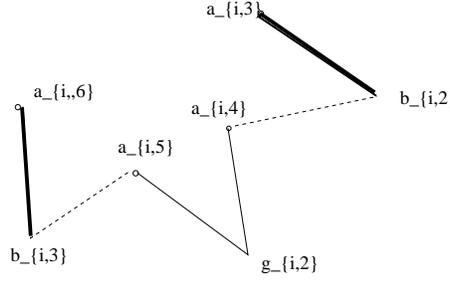}}
\caption{Thick lines indicate $M$-edges. If we match $g_{i,2}$ to $a_{i,4}$, we get a path $(a_{i,5},g_{i,2},a_{i,4},b_{i,2},a_{i,3})$, which is of type 
$(1)$.}
\label{turbinaprzyklad}
\end{figure}
Suppose there exists a weakly popular $b$-matching $M$. We will show that there exists a function $f: \{p_1, \ldots, p_k\} \rightarrow \{0,1\}$ such that $F$ is true when $p_i=f(p_i)$ ($1 \leq i \leq k$). The key observation is that
for any variable $p_i$ either all houses $p_{i,j}$ will be matched to agents $a_{i,j}$ or all houses $p'_{i,j}$
will be matched to agents $a_{i,j}$. We show this as follows.
We notice that in the subgraph corresponding to variable $p_i$ either all agents $a_{i,j}$ for even $j$ are matched
to houses $b_{i,j/2}$ or all agents $a_{i,j}$ for odd $j$ are matched to houses $b_{i,(j+1)/2}$. Figure \ref{turbinaprzyklad}
shows what happens when it is not the case. Then there exist two agents $a_{i,j}, a_{i,j+1}$ such that both are not matched via a rank two edge and the graph contains a path of type $2$, which means that $M$ is not weakly popular. 
Suppose that all agents  $a_{i,j}$ such that $j$ is even are matched via a rank two edge. Then they must also be
matched via a rank one edge. For suppose that $a_{i,2}$ is not matched to $p'_{i,1}$. Then $(a_{i,1}, b_{i,1}, a_{i,2},p'_{i,1})$ is either a beginning of a path of type $(1)$ or is a path of type $(3)$. 
Thus all houses $p'_{i,j}$ are matched to agents $a_{i,j}$. We set $f(p_i)=1$.
Next, let us consider clauses. Suppose the $1$st clause is of the form $(p_1 \vee not\ p_4 \vee p_7)$ and $c_{i,1}$ is matched to $h_i$. Then $c_{i,1}$ must also matched to some $p_{1,j}$, because otherwise $(c_{i,2}, h_i, c_{i,j}, p_{1,j})$ is a beginning of a path of type $1$ or $3$, which shows that the $1$st clause is satisfied under function $f$.
\koniec

\begin{theorem}
The problem of deciding whether a given triple $(G,b,r)$  admits a weakly popular $b$-matching is $NP$-hard, even if
 each applicant $a \in A$ has at most $3$ edges of rank $1$ incident with him/her and at most $1$ edge of rank $2$ and capacity at most $4$ and each house $h \in H$ has capacity $1$.
\end{theorem}
 The proof is by reducing the $3$-exact cover problem to the problem from the theorem.
 
\section{Polynomial algorithms}

We start with the case when there are no ties and each agent uses at most two ranks, houses have arbitrary capacities.

We are going to need an algoritm for the following problem. We have a bipartite graph $G=(A \cup H, E)$,$b:A \cup A \rightarrow N$ such that $b(a)=1$ for every $a \in A$, a partition
of $A$ into $A_1, A_2, \ldots, A_p$ and a nonnegative integer $k_i$ for each $1 \leq i \leq p$. We want to find a maximum $b$-matching $M$ of $G$ such that for each $1 \leq i \leq p$ the number of vertices of $A_i$ matched in $M$ is at least $k_i$ or ascertain that such a matching does not exist. Here we will use the term of a matching instead of a 
$b$-matching.

Let $M_{max}$ denote any maximum matching of $G$.
Clearly if a matching satisfying our requirements exists (further on we will call such a matching {\bf a partition matching}), $\Sigma_{i=1}^{p} k_i$ must not surpass $|M_{max}|$.
Suppose that matching $M$ is of maximum cardinality but it matches less than $k_i$ vertices of $A_i$. Then if the solution to our problem exists, there is in the graph a sequence $P_1, P_2, \ldots, P_s$ of edge-disjoint alternating paths such that
each $P_j=(a_j, \ldots, a'_{j})$ begins with a non-$M$-edge and ends with an $M$-edge, $a_1$ is of $A_i$ and is unmatched in $M$, $a_s$ is of $A_{i'}$ such that $M$ matches more than $k_{i'}$ vertices of $A_{i'}$ and for each $j \leq s-1$ vertices $a'_j$
and $a_{j+1}$ are of the same set $A_t$ for some $t$ and they either denote the same vertex or $a_{j+1}$ is unmatched in $M$. Let us call such a sequence
an {\bf improving sequence}. (To see that an improving sequence exists,  let $M'$ denote any partition
matching and consider $M \oplus M'$. Since in $M'$ at least $k_i$ vertices are matched there exists in $M \oplus M'$ an alternating path $P'_1$ beginning with an unmatched in $A_i$ vertex and ending on some vertex $a'_1 \notin A_i$. If $a'_1 \in A_t$ and $A_t$ is such that $M$ matches exactly $k_t$ vertices of $A_t$, then there exists an alternating path $P'_2$ (edge-disjoint from $P'_1$) beginning either at $a'_1$ or at a vertex $a_2 \in A_t$ unmatched in $M$ and ending on a vertex $a_3\ in A_s$ such that $t \neq s$ and $i \neq s$. Proceeding in this way we obtain an improving sequence.) The algorithm for this problem can be then described as follows. \\

\noindent{\scriptsize \underline{Algorithm Partition Matching} \\ \\
{\bf Input:} graph $G=(A \cup H, E)$, $b:A \cup H \rightarrow N$, a partition
of $A$ into $A_1, A_2, \ldots, A_p$, a sequence of nonnegative integers $(k_1, k_2, \ldots, k_p)$ \\
{\bf Output:} $b$-matching $M$ of maximum cardinality that for each $1 \leq i \leq p$ matches at least $k_i$ vertices of $A_i$,
or a report that such a matching does not exist. \\
\\
Find any maximum $b$-matching $M$ of $G$. \\
{\bf while} $M$ does not satisfy requirements: \\
\indent Find an improving sequence $P$. \\
\indent If $P$ does not exist write ''does not exist'' and halt. \\
\indent Otherwise $M:=M \oplus P$.} \\

We need the above algorithm for the following. Suppose agents $A_h=\{a_1, a_2, \ldots, a_k\}$ have all a rank two edge incident with house $h$ that has capacity $c<k$ (and has no rank one edges incident with it). Then we are able to match only $p$ agents of $A_h$ to $h$. If we match $a_2$ to $h$, then $a_2$ should also be matched to a rank one house, because otherwise for any $a_1$  not matched to $h$, path $(a_1,h,a_2, h')$ (where $h'$ is a rank one house for $a_2$)
forms a beginning of a path of type $1$ or a path of type $3$. Thus we should find such a $M_1$ matching among rank one 
edges that $p$ agents of $A_h$ are matched in $M_1$. \\

\noindent{\scriptsize {\underline{Algorithm A}\\
{\bf Input:} graph $G=(A \cup H, E)$, function $b: A \cup H \rightarrow N$, a partition of $E=E_1 \cup E_2$ \\
{\bf Output:} a weakly popular $b$-matching $M$ or a report that it does not exist \\
\\
Let $G_1=(A \cup H, E_1)$ and $b_1$ be defined as $b_1(a)=1$ for $a \in A$ and $b_1(h)=b(h)$.\\ 
Let $G_2=(A \cup H, E_2)$ and $b_2$ be defined as $b_2(a)=1$ for $a \in A$ and $b_2(h)=b(h)-deg_{E_1}(h)$ for $h \in H$. \\
If in $G_2$ every  $h \in H$ satisfies: $deg_{E_2}(h) \leq b_2(h)$, then \\
find a maximum $b_2$-matching $M_2$ of $G_2$, a max. $b_1$-matching $M_1$ of $G_1$  and output $M=M_1 \cup M_2$. \\
Otherwise let $H'=(h_1, h_2, \ldots, h_p)$ denote all houses $h \in H$ such that $deg_{E_2}(h) > b_2(h)$. For each $h_i \in H'$ build $A_i = N_{G_2}(h_i)$, where $N_{G_2}(h_i)$ denotes the set of neighbours of $h_i$ in $G_2$. Set $k_i=b_2(h_i)$. Set $A_{p+1}=A \setminus \bigcup_{1 \leq i \leq p} A_i$ and $k_{p+1}=0$.\\
Using algorithm Partition Matching for the input: $G_1, b_1$, the partition of $A$ into $A_1, \ldots, A_{p+1}$ and
$(k_1, \ldots, k_{p+1})$ compute a partition matching $M_1$. \\
If it does not exist write ''does not exist'' and halt. \\
Otherwise  for each $i, \ 1 \leq i \leq p$ let $A'_i \subseteq A_i$ denote the set of agents that are matched in $M_1$.
For each $a \in A'_i$ add $(a,h_i)$ to $M_2$. For each $a \in A_{p+1}$ add $(a,h)$, such that $(a,h) \in E_2$, to $M_2$. \\
Output $M=M_1 \cup M_2$.}}

\begin{figure} 
\centering{\includegraphics[scale=0.5]{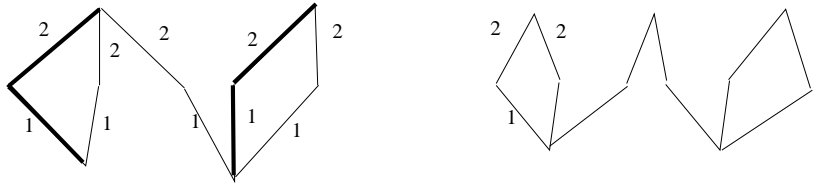}}
\caption{\scriptsize The graph on the left has a weakly popular $b$-matching and the one on the right has not.(Houses have capacity $1$.)}
\label{example}
\end{figure}

\begin{theorem}
Algorithm A solves the Weakly Popular $b$-Matching problem for the cases when $b(a)=2$ for each $a \in A$,
each agent uses $2$ ranks and there are no ties.
\end{theorem}
\dowod
Suppose that algorithm $A$ computes a $b$-matching $M$. To prove that it is weakly popular, it suffices by Theorem \ref{char1} to show that the graph does not contain a path of type $(1)$, $(3)$ or $(4)$.

Since $M$ is of maximum cardinality, the graph does not contain a path of type $3$ and since there are no ties,
the graph does not contain a path of type $4$. The graph does not contain a path of type $1$ either,
because, if such a path existed, then it would have the form $(a_1, h_1, a_2, h_2, a_3)$ such that $a_1$ is unsaturated
in $M$, edges $(a_1,h_1),(h_1, a_2)$ are of rank $2$ and edges $(a_2,h_2),(h_2,a_3)$ are of rank $1$. However $M$ has the property  that if agent $a$ is matched with a rank $2$ edge, then he/she is also matched with a rank $1$ edge.

On the other hand if algorithm A fails to compute a weakly popular $b$-matching, then it is because an appropriate partition matching does not exist. 
First let us notice that if a weakly popular $b$-matching exists, then it is of maximum cardinality.
Next we can show, that if in the graph $G$ a weakly popular $b$-matching $M$ exists, then there exists a weakly popular $b$ matching $M'$ such that $M'$ contains some maximum $b_1$-matching of $G_1$, where $b_1$ and $G_1$ are as in the description of
Algorithm A. For assume, that a weakly popular $b$-matching $M$ restricted to rank one edges (called $M_1$) is not a maximum $b_1$-matching of $G_1$. Then in $G_1$ there exists an $M_1$-augmenting path, which must be of the form $(a,h)$, where $a$ is not matched in $M_1$ and $h$ is not saturated in $M_1$. Since $M$ is weakly popular $h$ is saturated in $M$. Therefore there exists $a'$ such that $(a',h) \in M$ and $(a',h) \in E_2$. It is not difficult to see
that $M \setminus \{(a',h)\} \cup \{(a,h)\}$ must also be a weakly popular $b$-matching of $G$. We can proceed in this way until we have a weakly popular $b$-matching of $G$ that contains a maximum $b_1$-matching of $G_1$. 

Thus if there exists  a weakly popular $b$-matching $M$ of $G$, then there exists such a weakly popular $b$-matching $M'$ that $M'$ restricted to rank two edges is a maximum $b_2$-matching of $G_2$. If for some $h$ we have $deg_{E_2}(h) > b_2(h)$, then in any maximum $b_2$-matching of $G_2$ exactly $b_2(h)$ vertices of $N_{G_2}(h)$ will be matched and the remaining $deg_{E_2}(h) - b_2(h)$ vertices of $_{G_2}(h)$ will be unmatched. Therefore for such $h$ there will always be in $G$
a path $(a,h,a')$ such that $a,a' \in N_{G_2}(h)$ and $a$ is not matched in $G_2$. So if such a path is not to become
a beginning od a path of type $1$, each $a'$ that is matched in $G_2$ should also be matched in $G_1$. Therefore if there
exists a weakly popular $b$-matching of $G$, there exists an appropriate partition matching of $G_1$.
\koniec

Next we are going to deal with the case when ties are allowed among rank two edges and  each agent has capacity $2$ .
Suppose that a function $b: A \cup H \rightarrow N$ is such that $b(a)=1$ for each $a \in A$ and $M$ is a maximum
$b$-matching. Then a vertex $v \in A \cup H$ will be called an {\bf $O$-vertex} if there exists an odd-length alternating
path from un unmatched (in $M$) vertex $v_0$ to $v$ and an {\bf $E$-vertex} if there exists an  even-length alternating
path from un unmatched (in $M$) vertex $v_0$ to $v$. All other vertices will be called {\bf $U$-vertices} (unreachable via an alternating path from un unmatched vertex). By the Gallai-Edmonds decomposition theorem $O$-, $E$- and $U$-vertices forms a partition of $A \cup H$, which is independent of a particular maximum $b$-matching $M$. (See \cite{Lov}, \cite{rank} for example.)

We will need an algorithm for the following problem.
The input is as for Algorithm Partition Matching and additionally for each $i, \ 1 \leq i \leq p$ we have a family
$Z_i$ of subsets of $A_i$, each $Z \in Z_i$ having cardinality $k_i$.  We want to find a maximum matching $M$ of $G$ such that for each $1 \leq i \leq p$  vertices of $A_i$ matched in $M$ contain some $Z \in Z_i$ (such a matching
is going to be called a {\bf z-partition matching}) or ascertain that such a matching does not exist.
\begin{figure} 
\centering{\includegraphics[scale=0.5]{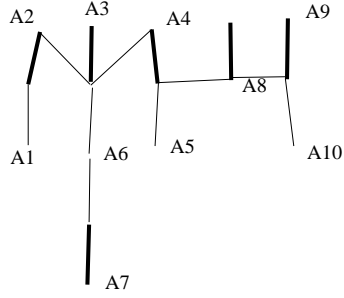}}
\caption{\scriptsize Thick lines indicate $M_2$-edges. Houses have capacity $1$. All agents except for $A_8$  are $O$-vertices. Agent $A_8$ is a $U$-vertex. Here two sets will belong to a partition of $A$: $\{A_1,A_2,A_3,A_4,A_5,A_6\}$ and $\{A_9, A_10\}$. $d(\{A_9, A_10\}=1$.}
\label{wazka}
\end{figure}

Let us first explain what we need this for. Suppose a maximum matching $M_2$ on some subgraph of $G_2$ looks as shown in Figure \ref{wazka}. Then if we want to avoid creating paths of type $(1)$,
we should have that agents $A_2, A_3, A_4, A_7, A_10$ have a rank one $M_1$-edge incident with them. Notice that
we do not have to worry about $A_8$.
If agents $Z_0=\{A_1, A_3, A_6, A_4, A_9\}$ have a rank one $M_1$-edge incident with them, then we can change $M_2$, so that
it will still be maximum and will saturate vertices of $Z_0$ and  the graph will not contain a path of type $1$.

With each subset $A_i$ of the partition of $A$ there will be associated a graph $G'_i=(A_i \cup H_i, E'_i)$ and a function
$b_2$ (such that $b(a)=1$ for each $a \in A$.)   We define $Z_i$ so that it contains each $Z \subseteq A_i$ such that there exists a maximum $b_2$-matching $M$ of
$G'_i$ having the property that vertices of $A_i$ matched in $M$ form the set $Z$. Notice that all sets in $Z_i$ have
the same cardinality.
For each $A'_i \subseteq A_i$ define $d(A'_i)=\max \{|B|: B \subseteq A'_i, \exists_{Z \in Z_i} B \subseteq Z\}$.
Suppose that in $A_i$, the set of matched in the current $b_1$-matching $M_1$ vertices equals $A'_i$. Then let $s_i=|A'_i|-d(A'_i)$.  
\begin{lemma}
For each $A'_i \subseteq A_i$, we can compute $d(A'_i)$ in polynomial time.
\end{lemma}
\dowod
It suffices to compute the largest subset $B \subseteq A$ such that there exists a maximum $b_2$-matching $M$ of $G'_i$, that
matches all vertices of $B$. To this end, first compute any maximum $b_2$-matching $M$ of $G'_i$. Let $B' \subseteq A'_i$ denote vertices matched in $M$. If $|B'|=|M|$ or $B'=A'_i$, set $B=B'$ and $d(A'_i)=|B|$.
Otherwise check if there exists in $G'_i$ an alternating path $P=(a_1, h_1, a_2, h_2, \ldots, a_n)$ such that
$a_1 \in A'_i \setminus B'$ and $a_n \notin A'_i$. If $P$ exists, set $M=M \oplus P$. This way new $M$ is clearly of maximum
cardinality (as $P$ is of even length) and $B'$ has increased by one. Repeat computing an alternating path of this type as long as possible. At the end $B'$ will be our desired set $B$.    
\koniec

If $A_i$ of the partition of $A$ is such that $s_i>0$ we will say that it is {\bf excessive} and if it such that
$d(A'_i)<|Z|$, where $Z \in Z_i$, we will say that it is {\bf deficient}. Notice that a set can be both excessive
and deficient. If a set is neither deficient nor excessive, then we call it {\bf equal}.

Suppose we have a maximum $b_1$-matching $M_1$ of $G_1$, but it is not a z-partition matching. Then in the partition
of $A$ there is at least one deficient set. We define a {\bf z-improving sequence} as a sequence $P_1, P_2, \ldots, P_s$ of edge-disjoint alternating paths such that
each $P_j=(a_j, \ldots, a'_{j})$ begins with a non-$M_1$-edge and ends with an $M_1$-edge, $a_1$ is from a deficient set, $a_s$ is from an excessive set, for each $j \leq s-1$ vertices $a'_j$
and $a_{j+1}$ are of the same equal set $A_t$ of the partition of $A$ (they can denote the same vertex)
and for each $j \leq s-1$ if $a'_j \in A_t$, then $A'_t \setminus \{a'_j\} \cup \{a_{j+1}\} \in Z_t$.  
\begin{lemma}
A z-improving sequence, if exists, can be found in polynomial time. 
\end{lemma}
\begin{lemma}
If z-improving sequence (with respect to any maximum $b_1$-matching of $G_1$) does not exist,
then  $G_1$ does not contain a z-partition matching. 
\end{lemma}

Algorithm for computing a z-partition matching looks exactly as algorithm for a partition matching, only it finds
a z-improving sequence instead of an improving sequence.

Algorithm A' is similar to Algorithm A, but it computes a different partition of $A'$. The partition in Algorithm A'
is established as follows. First we compute a maximum $b_2$-matching $M'_2$ of $G_2$. Next we find the $(E,O,U)$-partition
of $A$ into $O-,E-$ and $U$-vertices. Two $O$-vertices $a_1,a_2$ of $A$ will belong to one set of partition of $A$
iff there exists $h\in H$ such that there exists an $M_2'$-alternating path from $a_1$ to $h$ and from $a_2$ to $h$.
For such a set $A_i$ we will have $k_i= \sum_{a \in A_i} deg_{M'_2}(a)$. Additional set, say $(p+1)$th of the partition will be formed
by the remaining agents of $A$ and we will have $k_{p+1}=0$.

\end{document}